\documentstyle[11pt,LFnewpasp,twoside,epsf]{article} 
\def\edcomment#1{\iffalse\marginpar{\raggedright\sl#1\/}\else\relax\fi}
\reversemarginpar
\begin{document}
\title{The evolution of AGN in the Hard X-rays from HELLAS}
 \author{F. La Franca$^1$, F. Fiore$^2$, C. Vignali$^3$,
A. Comastri$^4$, F. Pompilio$^5$, plus HELLAS consortium}
\affil{
1) Dipartimento di Fisica, Universit\`a Roma Tre, Via della Vasca Navale 84, I-00146 Roma, Italy \\
2) Osservatorio Astronomico di Roma, Via Frascati 33, I-00044 Monteporzio Catone, Italy \\
3) Dipartimento di Astronomia, Via Ranzani 1, I-40127 Bologna, Italy \\
4) Osservatorio Astronomico di Bologna, Via Ranzani 1, I-40127 Bologna, Italy \\
5) SISSA, Via Beirut 4, I-34014 Trieste, Italy
}

\begin{abstract}
We present optical spectroscopic identification of sources identified
in the BeppoSAX MECS fields of the High Energy LLarge Area Survey
(HELLAS). In total 62 sources from a sample of 115 brighter than
$F_{5-10keV}>5\times10^{-14}$ $cgs$ have been identified. We find a
density of 13-21 sources/deg$^2$ at $F_{5-10keV}>5\times10^{-14}$
$cgs$, which contribute to $20-30$ \% of the hard X-ray background.
Evidences are found for type 1 AGN being more absorbed with increasing
redshift or luminosity. The low redshift ($z<0.2$) ratio of type 2 to
type 1 AGN is $3\pm1.5$, in agreement with the unified models for
AGN. The luminosity function of type 1 AGN in the 2-10 keV band is
preliminary fitted by a two-power law function evolving according to a
pure luminosity evolution model: $L\propto (1+z)^{2.2}$.
\end{abstract}

\section{Introduction}

Hard X-ray observations are the most efficient way of tracing emission
due to accretion mechanisms, such in Active Galactic Nuclei (AGN), and
sensitive hard X-ray surveys are powerful tools to select large
samples of AGN less biased against absorption and extinction.  In this
framework we decided to take advantage of the large field of view and
good sensitivity of the BeppoSAX MECS instrument (Boella et
al. 1997a,b) to survey tens to hundreds of square degrees at fluxes
$\ga 5-10\times10^{-14}~ erg~cm^{-2}s^{-1}$ (Fiore et al. 2000a), and
using higher sensitivity XMM-Newton and Chandra observations to extend
the survey down to $\sim10^{-14}~ erg~cm^{-2}s^{-1}$ on several
deg$^2$. The results from the optical identification of a sample of
faint Chandra sources discovered over the first 0.14 deg$^2$ have been
published by Fiore et al. (2000b). This approach is complementary to
deep pencil beam surveys ($\sim0.1$ deg$^2$, see e.g. Mushotzky et
al. 2000, Hornschemeier et al. 2000), as we cover a different portion
of the redshift--luminosity plane. Our purpose is to study cosmic
source populations at fluxes where
the X-ray flux is high enough to provide X-ray spectral information in
higher sensitivity follow-up observations. This would allow the
determination of the distribution of absorbing columns in the sources
making the hard XRB, providing strong constraints on AGN synthesis
models for the XRB (e.g. Comastri et al. 1995).

\section{The HELLAS survey}

The High Energy Large Area Survey (HELLAS, Fiore et al. 1999, 2000a,
2000c) has been carried out in the 4.5-10 keV band because: a) this
is the band closest to the maximum of the XRB energy density which is
reachable with the current imaging X-ray telescopes, and b) the
BeppoSAX MECS Point Spread Function (PSF) greatly improves with
energy, providing a 95\% error radius of 1$'$ in the hard band (Fiore
et al. 2000a), thus allowing optical identification of the X-ray
sources.

About 80 deg$^2$ of sky have been surveyed so far using 142 BeppoSAX
MECS fields at $|b|>18$ deg. 147 sources have been found with
$F(5-10keV)>5\times10^{-14}$ ${\rm erg~cm}^{-2}~{\rm s}^{-1}$ . We
find a density of 13-21 sources per square degree at
$F_{5-10keV}>5\times10^{-14}$ ${\rm erg~cm}^{-2}~{\rm s}^{-1}$, which
contribute to $20-30$ \% of the hard XRB in this energy range.

\begin{figure}[t]
\plotfiddle{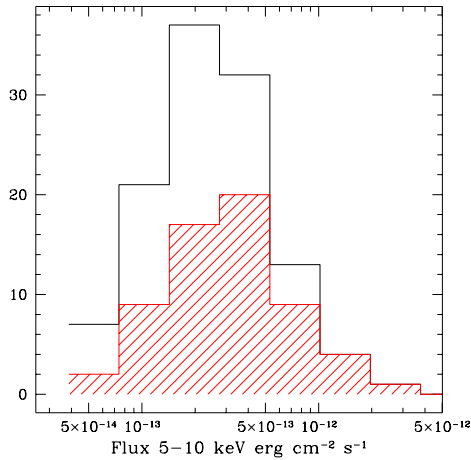}{180pt}{0}{35}{35}{-100}{-70}
\caption{The histograms of the flux distribution of the
subsample of 115 sources (continuous line), and of the 62
spectroscopic identified sources (dashed region). The two samples
have the same distribution at the 37$\%$ confidence level according to
the Kolmogorov-Smirnov test.}
\end{figure}

For the spectroscopical identifications we concentrated in a reduced
subsample of 110 MECS fields having $-60^\circ<\delta<+79^\circ$, and
excluding the regions with $5^h<\alpha<6.5^h$ and
$16.9^h<\alpha<20^h$. In this subsample the sky coverage is $\sim 55$
deg$^2$ at $5\times10^{-13}$ ${\rm erg~cm}^{-2}~{\rm s}^{-1}$ , and
0.5 deg$^2$ at $5\times10^{-14}$ ${\rm erg~cm}^{-2}~{\rm s}^{-1}$. We
found in total 115 such sources with $F(5-10keV)>5\times10^{-14}$ ${\rm
erg~cm}^{-2}~{\rm s}^{-1}$.

\section{Optical identifications}

Correlations of the HELLAS spectroscopic subsample of 115 sources with
catalogs of cosmic sources provide 26 coincidences (7 radio-loud AGN,
13 radio-quiet AGN, 6 clusters of galaxies). Optical spectroscopic
follow-ups have been performed on 49 of the 89 remaining HELLAS
error-boxes, providing 36 new identifications (Fiore et al. 1999, La
Franca et al. 2000 in preparation). We limited the optical
identification process to objects with surface density $<40$
deg$^{-2}$ to keep the number of spurious identifications in the whole
sample $<3-4$. As a consequence, we chose a limit of R$=20.5$ for
Broad line AGN; R$=19$ for Narrow line AGN (Sy1.8-2); and R$=17.5$ for
Emission line galaxies (LINERS and Starbursts). According to these
limits we found 13 ``empty'' fields in total. The summary of the
spectroscopic identifications is shown in Table 1.

\begin{center}
\begin{table}
\caption{The spectroscopic identifications}
\begin{tabular}{lccc}
\hline
 & total & new identifications & from catalogs \\ 
\hline
HELLAS			& 62 & 36 & 26 \\
Type 1 AGN	        & 36 & 21 & 15 \\
Type 1.8-2 AGN     	& 12 & 11 & 1  \\
Emission line galaxies	& 5  & 4  & 1  \\
BlLacs			& 2  & 0  & 2  \\
Radio galaxies		& 1  & 0  & 1  \\
Clusters of galaxies	& 6  & 0  & 6  \\
\hline
\end{tabular}
\end{table}
\end{center}

The correlation with existing catalogues did identify only 2 type 2
AGN and emission line galaxies. This is explained as there are no
existing large catalogues of such objects. Consequently a bias is
introduced in the observed fractions of the populations of the
identified sources against type 2 AGN and narrow emission line
galaxies. However the flux distribution of the sample of the 26
sources identified through cross-correlations with catalogues is
similar (according to the KS tests) to that of the total sample of 115
sources, and to the sample of 36 sources identified at the
telescopes. In this case we can correct for the introduced biases.
Applying these corrections we derive a value of 52$\%$ for the true
fraction of type 1 AGN in the total sample, and 29$\%$ for narrow
emission line galaxies and type 2 AGN. The remaining 19$\%$ are either
AGN with optical magnitudes fainter than our spectroscopic limits or
clusters of galaxies, BL-Lac etc. etc.

These numbers do not correctly address the question on the estimate of
the ratio of type 1 and 2 AGN. Our sample is also biased against high
redshift type 2 AGN which are indeed fainter than our spectroscopic
optical limit of $R$=19 for Narrow line AGN (Sy1.8-2), and R$=17.5$
for emission line galaxies. For this reason all the spectroscopic
identified type 2 AGN have $z<0.4$. A first preliminary estimate can
be drawn by taking into account only the AGN having $z<0.2$. In this
case we estimate that the ratio of the number of type 2 AGN and
emission line galaxies to the number of type 1 AGN is 3$\pm 1.5$, in
agreement with the unified models for AGN. More accurate analysis via
model predictions from the evolution of type 1 and 2 AGN are
underway.

In Figure 2 we plot the softness ratio (S-H)/(S+H) as a function of
the redshift for the 53 identified sources detected far from the
beryllium strong-back supporting the MECS window. Note that the
softness ratios of constant column density models strongly increases
with the redshift. Most of the narrow line AGN have (S-H)/(S+H)
inconsistent with that expected from a power law model with
$\alpha_E=0.4$. Absorbing columns, of the order of $10^{22.5-23.5}$
cm$^{-2}$, are most likely implied. Note also that some of the broad
line AGN have (S-H)/(S+H) inconsistent with that expected for a
$\alpha_E=0.8$ power law, in particular at high redshift. The
(S-H)/(S+H) of the 24 broad line AGN is marginally anti-correlated
with $z$ (Spearman rank correlation coefficient of -0.364, corresponding
to a probability of 92$\%$). The number of sources is not large enough
to reach a definite conclusion. Similar results have been recently
found in ASCA samples by Akiyama et al. (2000) and Della Ceca et
al. (2000). XMM-Newton and Chandra follow-up observations may easily
confirm or disregard a significant absorbing column in these high $z$
broad line quasars.

\begin{figure}[t]
\plotfiddle{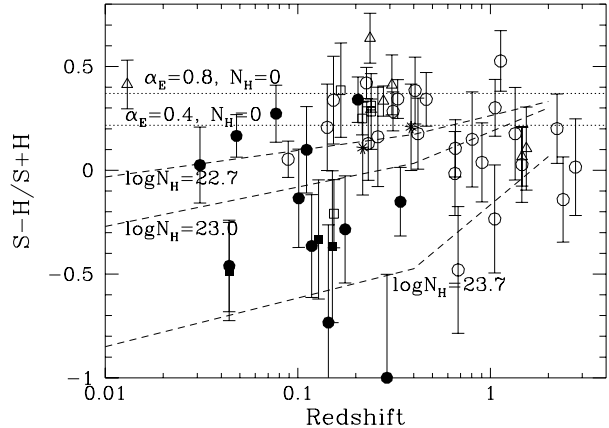}{180pt}{-90}{35}{35}{-140}{190}
\caption{The softness ratio
(S-H)/(S+H) versus the redshift for the identified sources.  open
circles = broad line, `blue' continuum quasars and Sy1; stars= broad
line `red' continuum quasars; filled circles= type 1.8-1.9-2.0 AGN;
filled squares= starburst galaxies and LINERS; open triangles=
radio-loud AGN; open squares= clusters of galaxies.  Dotted lines show
the expected softness ratio for a power law model with $\alpha_E$=0.4
(lower line) and $\alpha_E=0.8$ (upper line). Dashed lines show the
expectations of absorbed power law models (with $\alpha_E=0.8$ and
log$N_H$=23.7, 23.0, 22.7, from bottom to top) with the absorber at
the source redshift.  }
\end{figure}

\section{The evolution of AGN}

We combined our sample of hard X-ray selected AGN with other samples
of AGN identified by Grossan (1992), Boyle et al. (1998), and Akiyama
et al. (2000). A total of 151 type 1 AGN have been used in the redshift
range $0<z<3$. Values of $H_0=50$ Km s$^{-1}$ Mpc$^{-1}$, q$_0 = 0.5$,
$\Lambda = 0$ have been used.  The 2-10 keV luminosities have been
computed using the k-correction derived from the model of the spectrum
of type 1 AGN as in Pompilio, La Franca and Matt (2000). The coverage
of the $L-z$ plane by the whole sample of type 1 AGN is shown in
Figure 3a.

The sample of Grossan (1992) consists of 84 type 1 AGN and 12 type 2
AGN, predominantly at low redshift ($z<0.3$). The sample covers an
area of 26919 deg$^2$ down to a flux limit of $1.8\times
10^{-11}~cgs$.  The sample of Boyle et al. (1998) consists of 12 AGN1
and 6 AGN2.  The sample of Akiyama et al.\ (2000) consists of 25 AGN1,
5 AGN2, 2 clusters, 1 star and 1 unidentified source.

\begin{figure}[t]
\plottwo{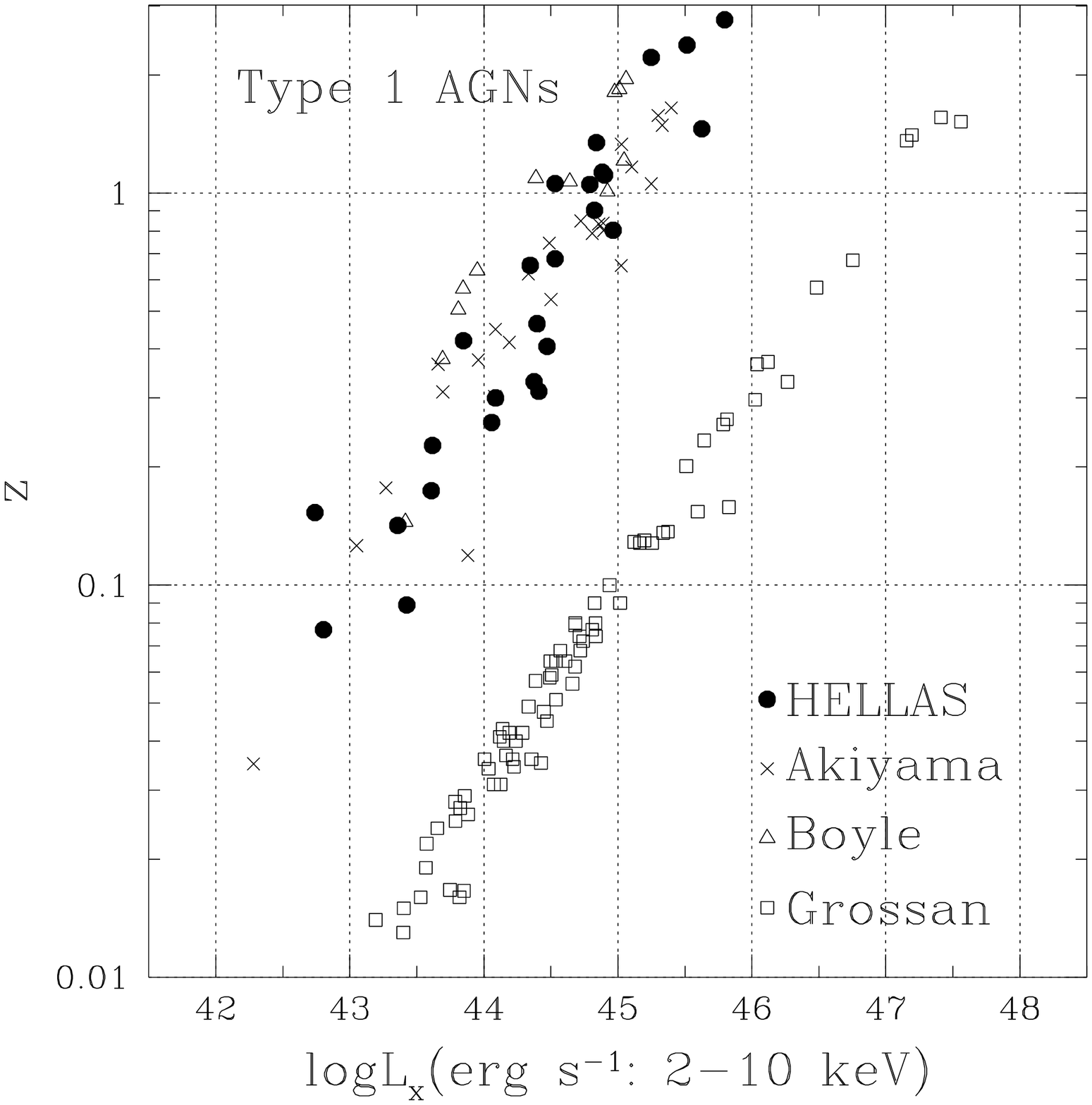}{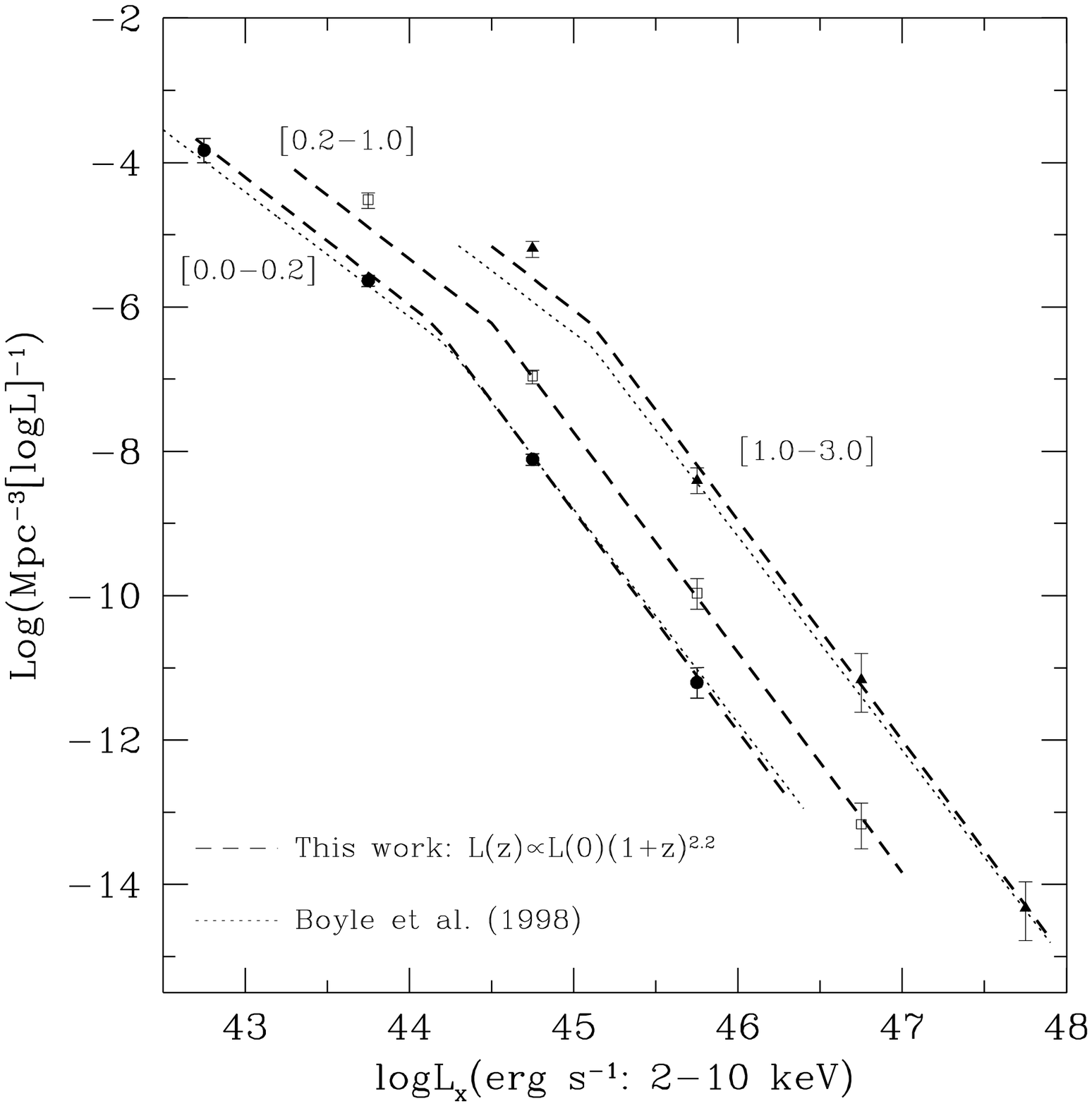}
\caption{a) The luminosity/redshift plane of the sample of type 1 AGN
used in the computation of the luminosity function. b) The luminosity
function of type 1 AGN as fitted by a pure luminosity evolution model
at redshifts 0.1, 0.6, and 2.0. The densities have been corrected for
evolution within the redshift bins.}
\end{figure}

We carried out a maximum likelihood analysis to derive best-fit
parameters which described the 2-10 keV luminosity function and its
cosmological evolution.  Following Boyle et al. (1988) and Ceballos
and Barcons (1996), we chose a two-power-law representation for the
QSO LF:

$$\Phi(L_X) = K_1 L_{X}^{-\gamma _1}:\ \ \ \ \ L_X < L_X^*(z=0)~$$
$$\Phi(L_X) = K_2 L_{X}^{-\gamma _2}:\ \ \ \ \ L_X > L_X^*(z=0).$$ 

\noindent
Where $K_1$ [$=K_2 (L_{X}^*/10^{44}{\rm erg\,s^{-1}}) ^{(\gamma _1 -
\gamma _2)}$].  A standard power-law luminosity evolution model was
used to parameterize the cosmological evolution of this luminosity
function: $ L_X^*(z) = L_X^*(0)(1+z)^{k}$.

A preliminary estimate of the fitted luminosity function is shown in
Figure 3b. The corresponding fitted values are $\gamma _1=1.8$,
$\gamma _2=3.1$, $k=2.2$, $logL_X^*(z=0)=44.1~erg~s^{-1}$, $K_2 =
8.8\times 10^{-7} Mpc^{-3}/(10^{44} erg~s^{-1})$. No redshift cut-off
in the luminosity evolution has been applied. This result is in good
agreement with the previous estimate from Boyle et al. (1998) on a
smaller sample of 95 type 1 AGN (see Figure 3b). The evolved
luminosity function shows a density excess of low luminosity high
redshift objects. A more detailed study of the luminosity function
which will take into account a possible redshift cut-off and density
evolution is underway.

\begin{acknowledgements}
We thank the BeppoSAX SDC, SOC and OCC teams for the successful
operation of the satellite and preliminary data reduction and
screening. This research has been partly supported by ASI 
ARS/99/75 contract and MURST Cofin-98-02-32 contract.
\end{acknowledgements}

%the {\it Chicago Manual of Style\/}.
%Journal names and book titles should be set in {\it italics\/}.
%Volume numbers should be {\bf boldface}.


\begin{references}

\reference
Akiyama, M., Ohta, K., Tamura, N., et. al., 2000, ApJ, 532, 700

\reference
Boella, G., Butler, R.C., Perola, G.C., et al., 1997a, \aaps, 122, 299 

\reference
Boella, G., Chiappetti, L., Conti, G., et al., 1997b, \aaps, 122, 327

\reference
Boyle, B.J., Georgantopoulos, I., Blair, A.J., Stewart G.C.,
Griffiths, R.E., Shanks, T., Gunn, K.F., Almaini. O., 1998,
\mnras, 296, 1

\reference
Ceballos, M.T., Barcons, X., 1996, \mnras, 282, 493

\reference
Comastri, A., Setti, G., Zamorani, G. \& Hasinger, G. 1995,
\aap, 296, 1

\reference
Della Ceca, R., Braito, V., Cagnoni, I., Maccacaro, T., 2000,
Proceedings of the Fourth Italian Conference on AGN, MemSAIt in press,
[astro-ph/0007431]

\reference
Fiore, F., La Franca, F., Giommi, P., et al., 1999, \mnras, 306, L55

\reference
Fiore, F., Giommi, P., Vignali, C., et al., 2000a, \mnras, submitted

\reference
Fiore, F., La Franca, F., Vignali, C., Comastri, A., Matt, G., Perola
G.C., Cappi, M., Elvis M., Nicastro, F., 2000b, New Astronomy, 5, 143

\reference
Fiore, F., et al., 2000c, Proceedings of "X-Ray Astronomy '99", Bologna (Italy),
[astro-ph/0007118]

\reference
Grossan, B.A., 1992, PhD thesis, MIT

\reference
Hornschemeier, A.E., Brandt, W.N., Garmire, G.P., et al., 2000, \apj, 541, 49

\reference
Mushotzky, R.F., Cowie, L.L., Barger, A.J., Arnaud, K.A.,
2000, Nature, 404, 459

\reference
Pompilio, F., La Franca, F., Matt, G., 2000, \aap, 353, 440

\end{references}
\end{document}